\documentclass[12pt]{article}
\newcommand{\eq}{\begin{equation}}
\newcommand{\feq}{\end{equation}}
\newcommand{\eqn}{\begin{eqnarray}}
\newcommand{\feqn}{\end{eqnarray}}
\newcommand{\arr}{\begin{eqnarray*}}
\newcommand{\farr}{\end{eqnarray*}}
\newcommand{\beq}{\begin{equation}}
\newcommand{\eeq}{\end{equation}}
\newcommand{\bea}{\begin{eqnarray}}
\newcommand{\eea}{\end{eqnarray}}

\newcommand{\TT}{{\cal T}}

\def\beq{\begin{equation}}
\def\eeq{\end{equation}}
\def\bea{\begin{eqnarray}}
\def\eea{\end{eqnarray}}
\def\bc{\begin{displaymath}}
\def\ec{\end{displaymath}}
\def\lb{\label}

\def\ha{{1\over2}}
\def\gr{\sqrt{-g}}

\def\rF{{\rm F}}

\def\al{\alpha}
\def\be{\beta}
\def\ga{\gamma}

\def\ep{\varepsilon}

\def\ka{\kappa}
\def\la{\lambda}

\def\si{\sigma}
\def\s{\sigma}
\def\t{\tau}
\def\om{\omega}

\def\r{\rho}

\def\epsi{\epsilon}

\def\lb{\label}

\begin{document}
\begin{titlepage}
\begin{flushright}
INFNCA-TH0204 \\
MIT-CTP-3272\\
\end{flushright}
\vspace{.3cm}
\begin{center}
\renewcommand{\thefootnote}{\fnsymbol{footnote}}
{\Large \bf Two-dimensional dS/CFT correspondence}
\vfill
{\large \bf {M.~Cadoni$^{1,a}$\footnote{email: mariano.cadoni@ca.infn.it},
P.~Carta$^{1,a}$\footnote{email: paolo.carta@ca.infn.it},
M. Cavagli\`a$^{2,b}$\footnote{email: cavaglia@mitlns.mit.edu}
and S.~Mignemi$^{3,a}$\footnote{email: smignemi@unica.it}}}\\
\renewcommand{\thefootnote}{\arabic{footnote}}
\setcounter{footnote}{0}
\vfill
{\small
$^1$ Universit\`a di Cagliari, Dipartimento di Fisica,\\
Cittadella Universitaria, 09042 Monserrato, Italy\\
\vspace*{0.4cm}
$^{2}$ Center for Theoretical Physics, Massachusetts Institute of Technology,
77 Massachusetts Avenue, Cambridge MA 02139-4307, USA\\
\vspace*{0.4cm}
$^3$ Universit\`a di Cagliari, Dipartimento di Matematica,\\
Viale Merello 92, 09123 Cagliari, Italy\\
\vspace*{0.4cm}
$^a$ INFN, Sezione di Cagliari\\
\vspace*{0.4cm}
$^b$ INFN, Sede di Presidenza, Roma}
\end{center}
\vfill
\centerline{\bf Abstract}
We investigate de Sitter/conformal field theory (dS/CFT) correspondence in
two dimensions. We define the conserved mass of de Sitter spacetime and
formulate the correspondence along the lines of anti-de Sitter/conformal
field theory duality. Asymptotic symmetry group, mass, and central charge
of de Sitter spacetime are equal to those of anti-de Sitter spacetime. The
entropy of two-dimensional de Sitter spacetime is evaluated by applying
Cardy formula. We calculate the boundary correlators induced by the
propagation of the dilaton in two-dimensional de Sitter space. Although
the dilaton is a tachyonic perturbation in the bulk, boundary conformal
correlators have positive dimension.
\vfill
\end{titlepage}
\section{Introduction}
Recently, Strominger proposed a correspondence between gravity on
$d$-dimensional de Sitter space and $(d-1)$-dimensional conformal field
theory \cite{Strominger:2001pn,Spradlin:2001pw}. Evidence of a positive
cosmological constant $\la$ provided by astrophysical observations
\cite{Church:2001uy,Albrecht:2001xp} suggests that we live in a de Sitter
spacetime. An important feature of de Sitter spacetime is the existence of
a cosmological horizon endowed with entropy \cite{Gibbons:mu}. De
Sitter/Conformal Field Theory (dS/CFT) correspondence may hold the key to
its microscopical interpretation. Moreover, new investigations have
revealed the existence of holographic cosmological bounds on entropy and a
correspondence between cosmological Friedmann equations and Cardy formula
of CFT \cite{Verlinde:2000wg,Cadoni:2002rr}. dS/CFT duality could be
crucial in the understanding of the holographic principle in cosmology.

Naively, we would expect dS/CFT correspondence to proceed along the lines of
Anti-de Sitter/Conformal Field Theory (AdS/CFT) correspondence because de
Sitter spacetime can be obtained from anti-de Sitter spacetime by analytically
continuing the cosmological constant to imaginary values. However, local and
global properties of de Sitter spacetime lead to unexpected obstructions.
Unlike anti-de Sitter, the boundary of de Sitter spacetime is spacelike and its
dual CFT is Euclidean. Moreover, de Sitter spacetime does not admit a global
timelike Killing vector. The time dependence of the spacetime metric precludes a
consistent definition of energy and the use of Cardy formula to compute de
Sitter entropy. Finally, dS/CFT duality leads to boundary operators with
complex conformal weights, i.e., to a non-unitary CFT. In spite of these
difficulties, some progress towards a consistent definition of dS/CFT
correspondence has been achieved. A new procedure \cite{Balasubramanian:2001nb}
for the computation of the boundary stress tensor allows the definition of a
conserved mass and the calculation of the entropy of asymptotically de Sitter
spacetimes \cite{Balasubramanian:2001nb,Ghezelbash:2001vs}. In
the three-dimensional case, by far the best-known example of the
dS$_{d}$/CFT$_{d-1}$ correspondence, the
central charge of the
dual CFT has been computed and used in the Cardy formula to evaluate the entropy
\cite{Strominger:2001pn,Klemm:2001ea,Maldacena:1998ih,Myung:2001ab}.

In this paper we investigate dS$_{d}$/CFT$_{d-1}$ correspondence in
two-dimensions. Previous investigations of dS$_{2}$/CFT$_1$ duality have
only considered the quantization of scalar fields in two-dimensional (2D)
de Sitter spacetime \cite{Ness:2002qr}. Here, we analyze the
dS$_{2}$/CFT$_1$ correspondence in a full dynamical context, i.e., with 2D
de Sitter spacetime emerging as a solution of the field equations. The
main obstruction to the implementation of dS$_{2}$/CFT$_1$ correspondence
along the lines of AdS$_{2}$/CFT$_1$ correspondence is the definition of a
conserved mass for de Sitter spacetime. We show that a procedure similar
to that of Ref.\ \cite{Balasubramanian:2001nb} enables the formulation of
dS$_{2}$/CFT$_{1}$ correspondence in analogy to the AdS$_{2}$/CFT$_{1}$
case \cite{Cadoni:1999ja}. The generators of the asymptotic symmetric
group of dS$_{2}$ satisfy a Virasoro algebra. We compute the central
charge of the algebra by adapting to dS$_{2}$/CFT$_1$ the canonical
formalism of AdS$_{2}$/CFT$_{1}$ correspondence \cite{Cadoni:1999ja} and
its interpretation as Casimir energy \cite{Cadoni:2000fq,Cadoni:2000kr}.
The entropy of 2D de Sitter spacetime is evaluated by applying Cardy
formula. In the second part of the paper we calculate the correlators
induced on the one-dimensional boundary of the spacetime by the
propagation of the dilaton in the 2D bulk. Although the dilaton is a
tachyonic perturbation in the 2D spacetime, the dual boundary operator has
positive conformal dimension $h=2$. This somehow unexpected result seems
to be a general feature of the dS/CFT correspondence.
\section{2D Cosmological solutions of de Sitter gravity }
Let us consider the 2D dilaton gravity model with action
\beq\lb{action}
I={1\over2}\int\gr\ d^2x\ \Phi\left(R-2\lambda^{2}\right)\,,
\eeq
where $\Phi$ is the dilaton field, $\la$ is the cosmological  constant, and $R$
is the 2D Ricci scalar. The general solution of the model (\ref{action}) 
describes  a 2D
hyperbolic manifold with constant positive curvature $R=2\la^{2}$ (de Sitter
spacetime) endowed with a non-constant dilaton.
Two-dimensional de Sitter space  can be defined as the hyperboloid
\beq\lb{hyp}
X^{2}+Y^{2}-Z^{2}={1\over\la^2}
\feq
embedded in the three-dimensional spacetime with hyperbolic metric $ds^{2}=
dX^{2}+dY^{2}-dZ^{2}$. De Sitter spacetime can be interpreted as the
analytical continuation $\la\to i \la$ of  anti-de Sitter
spacetime. De Sitter spacetime is geodesically complete. However, the presence
of the dilaton field leads to three globally nonequivalent solutions which are
described by coordinate charts covering different regions of the de Sitter hyperboloid.
In analogy to the AdS case \cite{Cadoni:1994uf}, we call these solutions
dS$_{0}$, dS$_{-}$, dS$_{+}$:
\vskip 1em
\noindent
{\bf dS}$\bf _{0}$. In conformal coordinates the general solution of 
the gravity model  (\ref{action}) is
\eq\lb{sol1}
ds^2 = {1\over \la^{2} \t^{2}} (-d\t^2 + dx^2)\,,\quad
\Phi={{\al(x^{2}-\t^{2}) +\be x +\gamma}\over \t}\,,
\feq
where $\al$, $\be$, and $\ga$ are integration constants. The spatial sections at
$\tau=$ constant are either the line ($-\infty<x<\infty$) or the one-dimensional
sphere $S^{1}$ ($-\pi<x<\pi$). The solution (\ref{sol1}) is singular at
$\tau=0$. Therefore, dS$_{0}$ covers half of the de Sitter hyperboloid (\ref{hyp}).
Setting $\la \t=e^{\la T}$ the metric in Eq.\ (\ref{sol1}) reads
\eq\lb{sol2}
ds^2 = -dT^2 + e^{-2\la T} dx^2\,,
\feq
where $-\infty<T<\infty$. Equation (\ref{sol2}) is the $d=2$ case of the
dS$_{d}$ solution in planar coordinates
\cite{Strominger:2001pn,Spradlin:2001pw}. An interesting feature of the
2D solution is that the sections at $T=$ constant may have either
the topology of the line or of the circle. In higher  dimensions only planar
topologies are allowed. The $(T,x)$ coordinate  system covers half of the
de Sitter hyperboloid.  The spatial section at $T=-\infty$ ($\t=0)$ and
$T=\infty$ ($\tau=\infty$) are the spacelike boundary $\cal{I^{-}}$ of the
spacetime and the cosmological future horizon, respectively. Alternatively, we
can cover the other half of the de Sitter hyperboloid by setting
$\la\t=-e^{-\la T}$. In this case $T=-\infty$ ($\t=-\infty)$ and  $T=\infty$
($\tau=0$) are the cosmological past horizon and the spacelike boundary
$\cal{I^{+}}$ of the spacetime. Setting $\la t=1/(\la \t)$ the line element in
Eq.\ (\ref{sol1}) becomes
\eq\lb{sol2a}
ds^2 = -{1\over\la^{2}t^{2}}dt^2 + \la^{2}t^{2} dx^2\,.
\feq
The dS$_{0}$ solution is the analytic continuation $\la\to i\la$
of the 2D anti-de Sitter AdS$_{0}$ solution \cite{Cadoni:1994uf}.
\vskip 1em
\noindent
{\bf dS}$\bf _{-}$. Setting
\eq\lb{rm}
\t={1\over a\lambda} e^{a\la\s}\sinh(a\la\hat\t)\,,\qquad
x={1\over a\lambda} e^{a\la\s} \cosh(a\la\hat\t)\,,
\feq
the metric in Eq.\ (\ref{sol1}) is cast in the dS$_-$ form
\beq\lb{sol1a}
ds^2 = {a^2\over \sinh^{2}(a\la\hat\t)} (-d\hat\t^2 + d\s^2)\,.
\feq
The $(\hat \t, \s)$ coordinates cover the region $x^{2}\ge\t^{2}$ of
dS$_{0}$.  The coordinate transformation (\ref{rm}) is analogous to the
coordinate transformation which relates Minkowski and Rindler spacetimes.
Curves of constant $\s$ are hyperbola in the $(\t,x)$ coordinate frame and
represent world lines of accelerated observers. Analogously to the AdS
case \cite{Cadoni:1994uf}, dS$_{-}$ spacetime can be interpreted as the
thermalization of dS$_{0}$ spacetime at temperature $T_H=a\la/2\pi$.
Defining
\beq\lb{coor2}
\cosh\la T={\rm cotanh}{(a\la \hat\t)}\,,
\feq
Eq.\ (\ref{sol1a}) reads:
\eq\lb{sol3}
ds^2 = -dT^2 + a^2 \sinh^2{\la T} d\s^2\,,
\feq
where the cosmological time $T$ is defined in the interval $-\infty<T<0$.
Equation (\ref{sol3}) corresponds to the spherical slicing of higher
dimensional de Sitter spacetime \cite{Strominger:2001pn,Spradlin:2001pw}.
Similarly to the planar slicing (\ref{sol2}), the coordinate
$\s$ can parametrize either a line or a circle. Setting $\la t=a\cosh\la T$,
Eq.\ (\ref{sol3}) is cast in the form
\eq\lb{sol3a}
ds^2 = -{1\over \la^{2}t^{2}-a^2}dt^2 + (\la^{2}t^{2}-a^2)d\sigma^2\,.
\feq
dS$_{-}$ can be interpreted as the analytic continuation, $\la\to
i\la$, of the 2D
AdS$_{+}$ solution \cite{Cadoni:1994uf}.
\vskip 1em
\noindent
{\bf dS}$\bf _{+}$. In conformal coordinates the dS$_{+}$ spacetime is
described by the line element
\beq\lb{sol4}
ds^2 = {a^{2}\over \cos^{2}(a\la\t)} (-d\t^2 + d\r^2)\,,
\feq
where $-\pi/2a\le\t\le\pi/2a$. Equation (\ref{sol4}) describes the whole
de Sitter hyperboloid (\ref{hyp}). The spacelike coordinate $\rho$ can be
either periodic or defined on the real line. Setting
\beq\lb{coor3}
\sinh\la T=\tan({a\la \t})\,,
\feq
Eq.\ (\ref{coor3}) is cast in the form
\eq\lb{sol5}
ds^2=-dT^2+a^{2}\cosh^2{\la T}\,d\r^2\,,
\feq
where $-\infty<T<\infty$. The spatial sections at $T=\pm\infty$ ($\t=\pm
{\pi/2})$ are $\cal{I^{+}}$ and $\cal{I^{-}}$, respectively. Finally,
setting $\la t=a\sinh\la T$ the dS$_{+}$ line element becomes
\beq\lb{sol6}
ds^2 = -{1\over a^{2}+\la^{2}t^{2}}\,dt^2 +
\left(a^{2}+\la^{2}t^{2}\right)d\rho^2\,.
\feq
The dS$_{+}$ solution is the analytic continuation, $\la\to i \la$, of the 2D
AdS$_{-}$ solution \cite{Cadoni:1994uf}
\section{Isometries of 2D de Sitter spacetime and conserved mass}
The isometry group of 2D de Sitter spacetime is $SL(2,R)$. In the
coordinate system of Eq.\ (\ref{sol1}) the isometry group of dS$_{0}$ is
described by the Killing vectors
\beq\lb{kill}
\xi_{0}=(\t,x),\quad \xi_{1}=(0,2), \quad \xi_{-1}=(\t x, {1\over 2}
(\t^{2}+x^{2}))\,.
\feq
The $SL(2,R)$ algebra is generated by the operators
\beq\lb{killm}
L_{0}= \t\partial_{\t} +x\partial_{x},\,\quad L_{1}=2\partial_{x}\,,\quad
L_{-1}= \t x\partial_{\t} +{1\over 2}(\t^{2}+x^{2}) \partial_{x}\,.
\feq
$L_{0}$, $L_{1}$ and $L_{-1}$ generate dilatations, translations in $x$
and special conformal transformations, respectively. Any independent
Killing vector defines an independent conserved charge. A crucial point is
to identify the Killing vector that defines the energy of the solutions.
Since we are dealing with time-dependent cosmological solutions, we do not
expect any conserved charge associated with a globally timelike Killing
vector. The Killing vectors $\xi_{1}$ and $\xi_{-1}$ are spacelike on the
whole de Sitter hyperboloid. The Killing vector $\xi_{0}$ is timelike
(spacelike) for $\t^{2}>x^{2}$ ($\t^{2}<x^{2}$). In particular, $\xi_{0}$
is spacelike on the boundaries ${\cal I^{\pm}}$. As was pointed out in
Ref.\ \cite{Strominger:2001pn} for de Sitter spacetime in $d>2$
dimensions, the absence of a timelike conserved charge on the spacetime
boundaries ${\cal I^{\pm}}$ represents a serious obstruction to the
implementation of the dS/CFT correspondence. For $d>2$ a solution to this
problem has been proposed in Ref.\ \cite{Balasubramanian:2001nb}, where
the conserved mass of dS$_{d}$ space is defined as an integral on the
surface $\cal{S}$ orthogonal to a Killing vector $\tilde\xi$ of the
boundary metric. This definition identifies the mass of the dS spacetime
with the conserved charge of the theory living in its boundary. However,
the procedure of Ref. \cite{Balasubramanian:2001nb} cannot be implemented
in the dS$_2$/CFT$_1$ context because in this case the surface $\cal{S}$
is a point.

Using the results of  Ref.\ \cite{Mann:1992yv}, it is straightforward to prove
that no timelike Killing vector exists on the spacetime boundary of dS$_2$.
Moreover, any Killing vector $\xi^\nu$ of the metric must also be a Killing
vector of the dilaton field, i.e., $\xi^\nu$ must be a solution of the 
scalar
Killing equation
\beq
\xi^{\nu}\partial_{\nu} \Phi=0\,.
\feq
Given $\xi^\nu$, the quantity
\beq\lb{mann2}
\TT_\mu=T_{\mu\nu}\xi^\nu\,,\qquad \nabla^\mu\TT_\mu=0\,,
\feq
defines the conserved charge $Q$ through the equation
$\TT_\mu=\ep_\mu^{\;\nu}\nabla_\nu Q$. In general, the
dilaton gravity model (\ref{action}) admits a Killing vector of
the form
\beq\lb{mann}
\hat\xi^{\nu}= F_{0} \epsilon^{\nu\mu}\partial_{\mu}\Phi\,,
\feq
where $F_{0}$ is an arbitrary constant. The conserved charge is
\beq\lb{charge}
Q={F_{0}\over 2}\left( -\la^2 \Phi^{2}-(\nabla\Phi)^{2}\right)\,.
\feq
Equation (\ref{charge}) is a local and covariant definition of the conserved
charge. Substituting the dS$_{0}$ solution (\ref{sol1}) in Eqs.\ (\ref{mann})
and (\ref{charge}) the Killing vector and charge read
\eqn\lb{kill1}
\hat\xi&=&F_{0}\left(2\al x\t +\be\t, \al(\t^{2}+x^{2})+\be x
+\ga\right)\,,\\
\lb{cha1}
Q&=&{F_{0}\over 2}\la^{2}(4\al\ga-\be^{2})\,,
\feqn
respectively. As expected, $\hat\xi$ is a linear combination of the three
Killing vectors of the metric (\ref{kill}). On the boundaries
$\cal{I^{\pm}}$ the norm of $\hat\xi$ satisfies $\t^{2}|\hat
\xi|^{2}=(A+2\ga)^{2}$, where $A=x(\al x+\be)$. Therefore, the Killing
vector $\hat\xi^\nu$ is spacelike on $\cal{I^{\pm}}$ for any point of the
moduli space. Moreover, there is no value of the parameters $\al$, $\be$,
and $\ga$ such that $\hat\xi$ is everywhere timelike.

Although our model does not admit any global timelike Killing vector, Eqs.\
(\ref{kill1}) and (\ref{cha1}) define a one-to-one map between moduli space of
the dilaton and symmetries and conserved charges. We can single out solutions
with a given conserved charge by choosing the subgroup of $SL(2,R)$ that leaves
the dilaton invariant. Solutions invariant under dilatations are obtained by
choosing $\al=\ga=0$. In this case the dilaton is
\beq
\Phi= \be {x\over \t}\,,
\feq
and the conserved charge under dilatations is $Q_{D}=-{1\over 2}
F_{0}(\lambda\be)^{2}$. We may also require that the dilaton depends only on
time by setting $\al=\be=0$. This singles out the $x$-translation generator
from the $SL(2,R)$ isometry group of dS$_{0}$.

Summarizing, we fix the charge $Q$ (up to a multiplicative constant) by
choosing a point in the dilaton moduli space and identify the mass $M$ of
the cosmological solution with $Q$ itself. This procedure is the 2D
analogue of that of Ref.\ \cite{Balasubramanian:2001nb}: In two dimensions
the Killing vector of the boundary metric is
$\tilde\xi\propto\partial_{x}$ and the surface $\cal{S}$ which is
orthogonal to $\tilde\xi$ is a point. The above procedure enables us to
calculate the mass of dS$_{0}$, dS$_{-}$ and dS$_{+}$ solutions. If we
impose that the dilaton depends only on time, and use $t$ as timelike
coordinate, $\Phi=\la^2\ga t=\Phi_0\la t$ for the three different
parametrizations (\ref{sol2a}), (\ref{sol3a}) and (\ref{sol6}) of
dS$_{2}$.  Using this equation in Eq. (\ref{charge}), we find $M=0$ and
\beq\lb{mass1}
M=\pm F_{0}{\la^{2}•\over2}\Phi_0^{2}•a^2\,,
\eeq
for dS$_{0}$ and dS$_{\pm}$, respectively. The mass (\ref{mass1}) is
defined up to the overall arbitrary constant $F_0$. The sign of $F_0$ can
be fixed by requiring $M$ to be positive for the dS$_{-}$ solution
(``stability'' condition). The absolute value of $F_{0}$ is determined by
requiring that $M$ coincides with the mass defined as a boundary integral
(see Sect.\ 5). Together, these two conditions fix $F_{0}=-1/(\la\Phi_0)$.  
With this choice the energy is positive, zero, and negative for dS$_{-}$,
dS$_{0}$, and dS$_{+}$, respectively. The Killing vector $\hat\xi$ is
\beq\lb{mann1}
\hat\xi=(0,-1)\,.
\feq
Translations in $x$ have opposite direction with respect to usual
definition. With the above normalization the charge $Q$ of dS$_{-}$ is
positive. The spacelike component of $\hat\xi$ and the stress
energy-tensor are negative. The stability condition could also be enforced
by keeping the usual definition of the Killing vector,
$\hat\xi=(0,1)$, and reversing the sign of the action (and then of the
stress-energy tensor $T_{\mu\nu}$). This arbitrariness indicates that $Q$
cannot be identified with the physical energy of the gravity theory 
in the 2D bulk.
\section{Asymptotic symmetries of 2D de Sitter spacetime}
Let us consider the 2D de Sitter solutions dS$_{0}$, dS$_{-}$ and
dS$_{+}$. In the coordinate chart $(t,r)$, where $r=x,\sigma,\rho$
respectively for dS$_{0}$, dS$_{-}$ and dS$_{+}$, the Killing vectors
generating the asymptotic symmetry group of the metric are
\eqn\lb{asym}
\xi^t&=&-\epsilon'(r) t+{\alpha^t(r)\over t}+O(t^{-2})\,,\\ \nonumber
\xi^r&=&\epsilon(r)+{1\over 2}{\epsilon''(r)\over\lambda^4
t^2}+ {\alpha^r(r)\over t^{4}}+ O(t^{-5})\,.
\feqn
The asymptotic form of the line element and of the dilaton which are invariant
under the asymptotic symmetry group are \footnote{Analogously to the AdS case,
the asymptotic form of the dilaton field is not invariant under the
transformations generated by Eqs.\ (\ref{asym}) but changes with a term of the
same order of the field itself.}
\eqn\lb{boundcond}
g_{tt}&=& -{1\over \lambda^{2}t^{2}}+\gamma_{tt}(r){1\over \lambda^{4}t^{4}}
+ O(t^{-5})\,,\nonumber\\ 
g_{rr}&=& { \lambda^{2}t^{2}}+\gamma_{rr}(r)
+ O(t^{-1})\,,\\ 
g_{tr}&=& {\gamma_{tr}(r)\over \lambda^{3}t^{3}}
+ O(t^{-4})\,, \nonumber\\ \nonumber
\Phi&=&\Phi_{0}\left[ \lambda t+ \rho(r) \lambda t+
\gamma_{\phi\phi}(r){1\over \lambda t}
+ O(t^{-2})\right]\,.\nonumber
\feqn
The asymptotic deformations of the fields transform as
\eqn\lb{deform}
\delta\rho&=&\rho'\epsilon-(1+\rho)\epsilon'\,,\nonumber\\ 
\delta\gamma_{\phi\phi}&=&\gamma_{\phi\phi}'\epsilon+\gamma_{\phi\phi}\epsilon'
+{\rho'\over 2\lambda^2}\epsilon''+\lambda^2(1+\rho)\alpha^t\,,\nonumber\\ 
\delta\gamma_{tt}&=&\gamma_{tt}'\epsilon+2\gamma_{tt}\epsilon'+
4\lambda^2\alpha^t\,,\\ 
\delta\gamma_{rr}&=&\gamma_{rr}'\epsilon+2\gamma_{rr}\epsilon'+
{\epsilon'''\over\lambda^2}+2\lambda^2\alpha^t\,,\nonumber\\ 
\delta\gamma_{tr}&=&\gamma_{tr}'\epsilon+3\gamma_{tr}\epsilon'
-(\gamma_{tt}+\gamma_{rr}){\epsilon''\over\lambda}-\lambda{\alpha^t}'-
4\lambda^5\alpha^r\,.\nonumber
\feqn
In analogy with the AdS/CFT correspondence, we can compute the
generators of the asymptotic symmetry group. We must distinguish
two cases, depending on whether the spacelike sections at $t=$ constant are the
one-dimensional sphere $S^{1}$ ($0\le r\le 2\pi/\lambda$) or the real line
$R$ ($-\infty< r< \infty$):
\vskip 1em
\noindent
$\bf 0\le r\le 2\pi/\lambda$. For compact spatial sections $\epsilon$
can be expanded in Fourier series in the interval $[0, 2\pi/\lambda]$,
\eq\lb{eps-expansion}
\epsilon(r)=\sum_{k=0}^{\infty}\left[
a_k\cos( \lambda kr)+b_k\sin(\lambda kr)\right]\,.
\feq
The generators of the group of asymptotic symmetries are defined by
\eq\lb{vira-def}
\xi=\sum_{k=0}^{\infty}\lambda\left[
a_kA_k+b_kB_k\right]\,,
\feq
where
\eqn\lb{vira}
A_k&=&\left[kt + O(t^{-1})\right] \sin(\lambda kr)\partial_t+
{1\over\lambda} \left(1-{k^2\over 2\lambda^2
t^2}+O(t^{-4})\right)\cos(\lambda kr)\partial_r\,,\\
B_k&=&\left[-kt + O(t^{-1})\right]\cos(\lambda kr) \partial_t+
{1\over\lambda}\left(1-{k^2\over 2\lambda^2
t^2}+O(t^{-4})\right)\sin(\lambda kr)\partial_r\,.
\feqn
The algebra of the $A_k$ and $B_k$ is
\eqn\lb{vira-algebra}
& &[A_k,A_l]=\ha(k-l)B_{k+l}+\ha(k+l)B_{k-l}\,,\nonumber\\
& &[B_k,B_l]=-\ha(k-l)B_{k+l}+\ha(k+l)B_{k-l}\,,\\
& &[A_k,B_l]=-\ha(k-l)A_{k+l}+\ha(k+l)A_{k-l}\,,\nonumber
\feqn
Defining the new generators $L_k=iA_k-B_k$ Eqs.\ (\ref{vira-algebra})
assume the standard form of a Virasoro algebra
\eq\lb{new-vira-algebra}
[L_k,L_l]=(k-l)L_{k+l} + {c\over 12}(k^3-k)\delta_{k+l}\,,
\feq
where we have taken into account the possibility of a central extension $c$.
\vskip 1em
\noindent
$\bf -\infty<r<\infty$. In this case $\epsilon$ is expanded in Laurent series
\eq\lb{eps-expansion1}
\epsilon(r)=\sum_{k=-\infty}^{+\infty}\
a_k(\la r)^{k}\,.
\feq
The generators of the  algebra are defined by
\eq\lb{vira-def1}
\xi=\sum_{k=-\infty}^{+\infty}\ \la a_k\hat L_{k}\,,
\feq
where
\eqn\lb{vira1}
\hat L_{k}= \left[-k (\la r)^{k-1}t+O(t^{-1})\right]\partial_t+
{1\over \la}\left[ (\la r)^{k}+ {k(k-1)\over 2\la^{2}t^{2}}
(\la r)^{k-2} +O(t^{-4})\right]\partial_r\,.
\feqn
The algebra is
\eq\lb{vira-algebra1}
[\hat L_k, \hat L_m]=(m-k)\hat L_{k+m-1}\,.
\feq
Defining the new generators $L_k=-\hat L_{k+1}$, Eq.\ (\ref{vira-algebra1}) is cast
in the standard form (\ref{new-vira-algebra}).
\section{Central charge}
To calculate the central charge we use a canonical realization of the
asymptotic symmetries. Since the boundary is spacelike, we parametrize
the metric as
\beq\lb{param}
ds^{2}= N^{2}dr^{2}- \Sigma^{2}\left(dt+N^{t}dr\right)^{2}\,.
\eeq
The 2D space is foliated along the spacelike coordinate $r$.
Therefore, the dynamical evolution is generated by the Killing vector
$\xi^{r}$. Owing to the normalization of $\xi^{r}$, Eq.\ (\ref{mann1}),    the
integration measure along $r$ acquires an overall minus sign. Up to boundary
terms the action becomes
\beq\lb{eaction1}
I=-\int dr dt\left[{1\over N} \left(\Sigma' -
{\partial\over\partial t}(N^{t}\Sigma)\right)
\left(\Phi' -N^{t}\dot\Phi\right)
+N\left(\Sigma^{-1}\ddot\Phi-\Sigma^{-2}\dot\Sigma\dot\Phi
-\lambda^{2}\Sigma\Phi\right)\right]\,,
\eeq
where prime and dot denote differentiation with respect to $r$ and $t$,
respectively. Introducing the conjugated momenta
\beq
\Pi_{\Phi}={ \delta{\cal L}\over \delta \Phi'}\,,\qquad
\Pi_{\Sigma}={ \delta{\cal L}\over \delta \Sigma'}\,,
\eeq
the action reads
\beq\lb{eaction1a}
I=\int dr dt\left[\Pi_{\Sigma}\Sigma' +\Pi_{\Phi}\Phi'-N^{t}H^{t}
-NH^{r}\right]\,,
\eeq
where
\eqn\lb{h}
H^{t}&=&\Pi_{\Phi}\dot\Phi- \dot\Pi_{\Sigma}\Sigma\,,\nonumber\\
H^{r}&=&-\Pi_{\Phi} \Pi_{\Sigma}
+\Sigma^{-1}\ddot\Phi-\Sigma^{-2}\dot\Sigma\dot\Phi
-\lambda^{2}\Sigma\Phi\,.
\feqn
In order to have well-defined functional derivatives the Hamiltonian must be
supplemented by the surface term $J$:
\beq
{\cal H}= \int dt \left(N^{t}H^{t}+ NH^{r}\right) + J\,,
\feq
where
\beq\lb{var}
\delta J= \lim_{t\to\infty} \left[N(\Sigma^{-2}\dot\Phi\delta\Sigma
-\Sigma^{-1}\delta\dot\Phi)+ \dot N(\Sigma^{-1}\delta\Phi)
- N^{t}(\Pi_{\Phi}\delta\Phi-\Sigma\delta\Pi_{\Sigma})\right]\,.
\feq
In Eq.\ (\ref{var}) we have considered only the contribution of the
$t\to\infty$ boundary. The contribution of the $t=0$ boundary gives a similar
contribution. We will come back later to this point.

Let us calculate the conserved charge $Q$ which is associated with the Killing
vector $\partial_{r}$. The charge $Q$ will be identified with the mass of the
solution. For the dS$_{-}$ solution (\ref{sol3a}), we have $\delta J=\delta Q=
-(\la/2)\Phi_{0}\delta\Sigma^{-2}$. It follows
\beq\lb{mass}
Q={\la\over 2}\Phi_0a^{2}\,,
\feq
in agreement with Eq.\ (\ref{mass1}). The mass of the dS$_{+}$ solution
(\ref{sol6}) is $Q=-(\la/2)\Phi_{0}a^{2}$. 

Using Eq.\ (\ref{asym}) and Eq.\ (\ref{boundcond}) in Eq.\ (\ref{var}),
the variations of the charges $J(\ep)$ corresponding to the symmetries
generated by the Killing vectors (\ref{asym}) are
\beq\lb{var1}
\delta J(\epsilon)=-\Phi_{0}\left( \epsilon\la \left(\gamma_{rr}\delta
\rho- 2 \delta \gamma_{\phi\phi}+{1+\rho\over 2} \delta
\gamma_{tt}\right) +{1\over \la}\left(\epsilon''\delta \rho -
\epsilon'\delta \rho'\right)\right)\,.
\feq
The central charge $c(\epsi,\om)$ can be calculated from the deformation
algebra
\beq\lb{def}
\delta_{\om}J(\epsi)= \{J(\epsi),J(\om)\}_{DB}=J([\epsi,\om])+
c(\epsi,\om)\,.
\feq
Substituting Eq.\ (\ref{var1}) in Eq.\ (\ref{def}), and evaluating the
equation on the dS$_{0}$ background solution ($\rho=\gamma_{rr}=
\gamma_{tt}=\gamma_{\phi\phi}=0$ identically), we find
\beq\lb{cc}
c(\epsi,\om)={\Phi_{0}\over \la}\left(\epsilon''\om'-\epsilon'\om'' \right)\,,
\feq
where we have used Eq.\ (\ref{deform}). Analogously to the 2D AdS
case, the orthogonality problem \cite{Carlip:1999cy,Cadoni:1999ja} can be
solved by introducing the integrated charges (in this section we consider only
$r$ periodic)
\beq\lb{ic}
\hat J(\epsi)={\la\over2\pi}\int_{0}^{2\pi/\la}drJ(\epsi)\,.
\feq
The algebra (\ref{vira-algebra}) has central extension
\beq\lb{cc0}
c(A_{k},A_{l})= c(B_{k},B_{l})=0\,,\qquad c(A_{k},B_{l})= \Phi_{0}k^{2}l
\delta_{|k||l|}\,.
\feq
The central charge of the Virasoro algebra (\ref{new-vira-algebra}) is found by
shifting the $L_{0}$ operator by a constant. The result is
\beq\lb{cc1}
c=24\Phi_{0}\,.
\feq
The central charge of de Sitter is positive and equal to that of anti-de
Sitter. Following Ref.\ \cite{Cadoni:2000ah}, we can integrate locally the
variation (\ref{var1}) near the dS$_{0}$ background solution:
$J(\epsi)=-\Phi_{0}(\epsi''\rho -\epsi'\rho')/\la$. Since $J$ is defined up to
a total $r$-derivative, it follows
\beq
J(\epsi)= -{2\Phi_{0}\over \la} \epsi \rho''= \epsi\Theta_{rr}\,,
\feq
where $\Theta_{rr}$ can be identified as the stress energy tensor of the
one-dimensional boundary CFT. Using the transformation law of the boundary
field $\rho$ we verify that $\Theta_{rr}$ transforms as a stress-energy tensor
with central charge (\ref{cc1}).

Up to now we have considered only the contribution of the boundary at
$t=\infty$ . By taking into account the contribution of the boundary at $t=0$
(see, e.g., Ref.\ \cite{Cadoni:2000gm}) the total central charge is
\beq\lb{cc2}
c=12\Phi_{0}\,.
\feq
The result above can also be obtained by interpreting the central charge as
Casimir energy. (This method was first used in Ref.\ \cite{Cadoni:2000kr} for
2D AdS/CFT correspondence and subsequently in Ref.\
\cite{Klemm:2001ea} to calculate the central charge of three-dimensional de
Sitter spacetime.) The dS$_{-}$ line element (\ref{sol3a}) is related to the
dS$_{0}$ line element (\ref{sol2}) by the coordinate transformation
\eqn\lb{tra}
e^{\la T}={e^{a\la \s}\over \sqrt{\la^{2}t^{2}-a^{2}}}\,,\qquad
x={te^{a\la \s}\over a \sqrt{\la^{2}t^{2}-a^{2}}}\,.
\feqn
On the $t\to\infty$ boundary the coordinate transformation $x\to\sigma$ is
\beq\lb{pc}
x={e^{a\la \s}\over a \la}\,.
\feq
Equation (\ref{pc}) is the one-dimensional analogue of the plane-cylinder
transformation of a 2D conformal field theory. The stress-energy
tensor $\Theta_{xx}$ acquires a term which is proportional to the central
charge of the CFT and can be interpreted as a Casimir energy:
\beq\lb{tenstra}
\Theta_{xx}=\left(dx\over d\s\right)^{2}\Theta_{\s\s}-
{c\over 12}\left(dx\over d\s\right)^{2}\left\{\s,x\right\}\,,
\feq
where $\left\{\s,x\right\}$ is the Schwarzian derivative. Substituting Eq.\
(\ref{pc}) in Eq.\ (\ref{tenstra}), and recalling that $\Theta_{\s\s}=-\la M=0$
($\Theta_{xx}=-\la M= -{1\over2} \Phi_{0} a^{2}\la^{2}$) for dS$_{0}$
(dS$_{-}$), we obtain the result (\ref{cc2}).
\section{Entropy of de Sitter spacetime}
The solution (\ref{sol3a}) can be continued across the horizon $t^2<a^2/\la^2$:
\beq\lb{solbh}
ds^2=-(a^2-\la^2t^2)d\s^2+{dt^2\over a^2-\la^2t^2}\,,\qquad \Phi=\Phi_0\la t\,.
\eeq
Inside the horizon $t$ ($\si$) is spacelike (timelike). The metric is regular
and admits the timelike Killing vector $\xi_0={\partial\over\partial\s}$. The
dilaton has a naked timelike singularity at $t=0$, which is the
lower-dimensional analogue of the conical singularity of three-dimensional de
Sitter spacetime.

The temperature $T_{H}$ is the inverse of the period $\be$ that must be
assigned to the radial coordinate to avoid the conical singularity in the
Euclidean section, i.e.,
\beq\lb{TH}
T_{H}={1\over2\pi}{dg_{\s\s}\over dt}\Big|_{\rm hor}={\la a\over2\pi}\,.
\eeq
The entropy can be calculated by using the Lorentzian action (see Ref.\
\cite{Ghezelbash:2001vs}). The Euclidean formalism of Gibbons and Hawking
\cite{Gibbons:mu} is not suitable because the Euclidean action vanishes
identically. The bulk term is identically zero on the field equations,
and the Euclidean dS$_2$ (two-sphere) has no boundary contribution.  The
Lorentzian action $I$ is \cite{Yo}
\beq\lb{action-entropy}
I={1\over2}\int_M\sqrt{-g}\,d^2x\,\Phi(R-2\la^2)+\int_{\partial M}
\sqrt h\,d\si\,\Phi(K-K_0)\,,
\eeq
where $h$ is the metric at the boundary, $K=-{\dot g_{\si\si}\over2
\sqrt{g_{\si\si}}}$ is the trace of the extrinsic curvature evaluated at the
boundary $t\to\infty$, and $K_0$ is the trace of the extrinsic curvature
relative to the background metric dS$_0$. The unusual sign in front of the
boundary integral is due to the choice of normalization (\ref {mann1}).
Computing Eq.\ (\ref{action-entropy}) on the solution (\ref{sol3a}), the
boundary term gives $I=-(\be\Phi_0\la a^2)/2=-\pi\Phi_0 a$. By analytically
continuing the Gibbs-Duhem relation, we find
\beq\lb{entropy}
S=\be M-I=2\pi\Phi_0 a=2\pi\Phi_h\,,
\eeq
where $\Phi_h$ is the value of the dilaton at the horizon. Equation
(\ref{entropy}) is consistent with the thermodynamical relation $T_{H}=\partial
M/\partial S$. The entropy can also be computed by applying Cardy formula
to the boundary conformal field theory with central charge (\ref{cc2}):
\beq\lb{entropy2}
S=2\pi\sqrt{cl_0\over6}=2\pi\Phi_0 a\,.
\eeq
Equation (\ref{entropy2}) is in agreement with the semiclassical result
(\ref{entropy}).
\section{Boundary correlators}
In this section we discuss dS$_{2}$/CFT$_{1}$ correspondence by computing
correlation functions on the spacetime bulk and on its boundary. In
higher-dimensional de Sitter spacetimes this program is accomplished by
studying correlation functions of dual boundary operators induced by an
external field. The 2D model (\ref{action}), thanks to the
presence of a scalar degree of freedom (the dilaton field $\Phi$), enables to
compute correlation functions which are induced on the boundary by the
gravitational degrees of freedom of the bulk. Let us consider the field
equations of the dilaton
\beq\lb{dila}
\nabla_{\mu}\nabla_{\nu}\Phi= -\lambda^{2}g_{\mu\nu}\Phi.
\feq
2D dilaton gravity has no propagating physical degrees of freedom:
If we restrict ourselves to classical configurations, and fix the
diffeomorphism invariance of the theory, the dilaton does not propagate.
However, we allow dilaton deformations on the one-dimensional boundary of
dS$_{2}$ ($\rho$ and $\gamma_{\phi\phi}$ fields in Eq.\ (\ref{boundcond})).
These deformations correspond to pure gauge and off-shell dilaton propagation
on the spacetime boundary. Therefore, we require that the dilaton 
satisfies the trace equation
\beq\lb{dila1}
\nabla^{2}\Phi= -2\lambda^{2}\Phi\,,
\feq
instead of the full equations of motion (\ref{dila}). Equation (\ref{dila1})
is the equation of motion of a scalar field with negative mass-squared
$m^{2}=-2\lambda^{2}$ and describes the propagation of a tachyonic scalar field
in 2D de Sitter spacetime. Consider Eqs.\ (\ref{sol2}),
(\ref{sol3}) and (\ref{sol5}), where $r=x$ $,\sigma$, and $\rho$, respectively.
Owing to the presence of the cosmological horizon, no correlators between $r\in
\cal I^{-}$ and $r'\in \cal I^{+}$ exist for dS$_{0}$ and  dS$_{-}$. On the
contrary, dS$_{+}$ covers the whole de Sitter spacetime and non-trivial
correlators between $r\in \cal I^{-}$ and $r'\in \cal I^{+}$ exist. Let us deal
with the three cases separately:
\vskip 1em
\noindent
{\bf dS}$\bf _{0}$. In the background (\ref{sol2}) Eq.\ (\ref{dila1}) reads
\beq\label{fe1}
\left( -\partial_{T}^{2}+\la \partial_{T}+e^{2\la
T}\partial_{r}^{2}\right)\Phi=-2\lambda^{2}\Phi\,.
\feq
When $T\to -\infty$, the third term on the left hand side of Eq.\
(\ref{fe1}) is negligible. On the $\cal I^{-}$ boundary the dilaton is
\beq\lb{dilal}
\Phi\sim\phi_{-1}(r) e^{-\la T}\,.
\feq
Subleading terms can be evaluated by expanding $\Phi$ in powers of $e^{\la T}$:
\beq\lb{exp}
\Phi=\sum_{n=-1}^{\infty}\phi_{n}(r) e^{n\la T}\,.
\feq
Substituting Eq.\ (\ref{exp}) in Eq.\ (\ref{fe1}) we find
\beq\lb{dilaton1}
\Phi=\phi_{-1}(r) e^{-\la T}+\phi_{1}(r)e^{\la T}+O(e^{2\la T})\,,
\feq
where the index denotes the conformal dimension of the fields $\phi$.
dS$_0$ has no boundary fields with conformal dimensions $h=0$. The
conformal weights of subleading and leading terms in Eq.\ (\ref{dilaton1})
are consistent with the conformal transformation laws of $\rho$ and
$\gamma_{\phi\phi}$ with weights $h=\pm 1$, respectively. It is
interesting to compare the boundary condition (\ref{dilaton1}) with that
of a generic scalar field of mass $m$ that propagates on 2D de Sitter
spacetime \cite{Ness:2002qr}: $\Phi\sim e^{h_\pm\la T}$, where
$h_{\pm}=(1\pm\sqrt{1-4m^{2}/\la^{2}})/2$. Setting $m^{2}= -2\la^{2}$ we
find $h_{+}=-1$ and $h_{-}=2$. However, the previous result is only valid
for scalar fields with positive squared mass. For tachyonic fields we have
$h_{+}>1$ and the term of weight $h_{+}$ is subleading with respect to
$\phi_1$.

Changing the sign in the exponents of Eqs.\ (\ref{sol2}), (\ref{fe1}) and
(\ref{dilal}) we obtain the behavior of the dilaton on the boundary
$\cal{I^{+}}$:
\beq\lb{dilaton2}
\Phi=\phi_{1}(r) e^{\la T}+\phi_{-1}(r)e^{-\la T}+O(e^{- 2\la T})\,,
\feq
Leading and subleading terms of the dilaton on $\cal I^{+}$ are
interchanged with respect to $\cal I^{-}$. Generalizing to two-dimensions
the dS/CFT proposal of Ref.\ \cite{Strominger:2001pn}, the two-point
correlator of an operator $\cal O_{\phi}$ on $\cal I^{-}$ is derived from
the expression
\beq\lb{int}
{\cal J}=\lim_{T\to-\infty}\int_{\cal I^{-}} drdr'\left[e^{-\la(T+T')}\Phi(T,r)
\stackrel{\leftrightarrow}{\partial}_{T}G(T,r,T',r')
\stackrel{\leftrightarrow}{\partial}_{T'}\Phi(T',r')\right]_{T=T'}\,,
\feq
where $G$ is the de Sitter invariant Green function. (See Appendix.) Using
Eqs.\ (\ref{Pinf}), (\ref{had5}) and (\ref{dilaton1}) in Eq. (\ref{int}), we
find
\beq\lb{int1}
{\cal J}=\ka_{0}\int_{\cal I^{-}} drdr'\phi_{-1}(r)
\phi_{-1}(r'){1\over(r-r')^{4}}\,,
\feq
where $\ka_0$ is a constant. The two-point correlator of an operator ${\cal
O}_{\phi}$ dual to $\phi_{-1}$ is the coefficient of the quadratic term in Eq.
(\ref{int1}):
\beq\lb{corr1}
\langle{\cal O}_{\phi}(r){\cal O}_{\phi}(r')\rangle= {\ka_0' \over 
(r-r')^{4}}\,.
\feq
Equation (\ref{corr1}) is the two-point correlator of a conformal operator of
dimension $h=2$. The two-point correlator on $\cal I^{+}$ can be computed in a
similar way. The relevant boundary field is $\phi_{1}(r)$ and the dual operator
on $\cal I^{+}$ satisfies Eq.\ (\ref{corr1}). The previous results show that a
tachyonic perturbation of the bulk corresponds to a boundary operator of
positive conformal dimension. This feature is a consequence of the holographic
correspondence between gravity on the 2D bulk and CFT on the
boundary. Technically, the result follows from a general property of the
integral in Eq.\ (\ref{int}). The dual operator of a boundary field with
conformal dimension $h_{-}$ has dimension $h_{+}$. Therefore, the tachyonic
perturbation ($h_{-}<0$) is in correspondence with a boundary operator of
positive conformal dimension ($h_{+}>0$). This property seems to be a general
feature of the integral (\ref{int}) and we expect it to hold for dS/CFT duality
in any dimension.
\vskip 1em
\noindent
{\bf dS}$\bf _{-}$. In this case de Sitter spacetime is described by the metric
(\ref{sol3}). Equation (\ref{dila1}) is
\beq\label{fe2}
\left( -\partial_{T}^{2}-\la \coth(\la T) \partial_{T}+
{1\over a^{2}\sinh^{2}(\la T)} \partial_{r}^{2}\right)\Phi=-2\lambda^{2}\Phi\,.
\feq
Setting $T\to\pm \infty$ in Eq.\ (\ref{fe2}), we find that the asymptotic
behavior of the dilaton is given by Eqs.\ (\ref{dilaton1}) and
(\ref{dilaton2}) for $\cal{I^{-}}$ and $\cal{I^{+}}$, respectively. The
dS$_{-}$ boundary correlators are computed by substituting Eqs.\
(\ref{dilaton1}) and (\ref{dilaton2}) and the asymptotic expression of
$G$ and $P$ given in the Appendix in the integral (\ref{int}):
\beq\lb{corr2}
\langle{\cal O}_{\phi}(r){\cal O}_{\phi}(r')\rangle= {\ka_{-} \over
\sinh^{4}{a\la\over 2} (r-r')}\,.
\feq
Since dS$_{0}$ and dS$_{-}$ are locally identical, the  correlators
 (\ref{corr1}) and (\ref{corr2}) 
have the same $\Delta r= r-r'\to 0$, short-distance, behavior.
The global features of the spacetime become manifest at large $\Delta r$.
The sinh behavior in Eq.\ (\ref{corr2}) describes a thermal CFT with
temperature equal to the Hawking temperature of the cosmological horizon
of dS$_{-}$. This result can be explained in CFT as follows. Equation
(\ref{pc}) maps the boundary of dS$_{0}$ on the boundary of dS$_{-}$. This
transformation can be interpreted as the one-dimensional analogue of
the plane-cylinder map $\la
z=\exp(\la w)$ of a 2D CFT, where $w=-\bar w= -ir$. In complex coordinates
Eq.\ (\ref{corr2}) becomes
\beq\lb{corr3}
\langle{\cal O}_{\phi}(r){\cal O}_{\phi}(r')\rangle= {\ka_{-}' \over
[\sin(\pi T_{H}^í\Delta w)\sin(\pi T_{H}^í\Delta \bar w)]^{2}}\,,
\feq
where $\Delta w=w-w'$ and $T_{H}$ is the Hawking temperature of the
cosmological horizon (\ref{TH}). The appearance of thermal correlators can also
be understood in terms of the 2D gravity theory:  dS$_{-}$ can be
considered as the thermalization of dS$_{0}$ at temperature $T_{H}={\la a/
2\pi}$ (see Sect.\ 2).
\vskip 1em
\noindent
{\bf dS}$\bf _{+}$. In the background (\ref{sol5}) the equation of motion of
the dilaton is
\beq\label{fe3}
\left( -\partial_{T}^{2}-\la \tanh(\la T) \partial_{T}+
{1\over a^2\cosh^{2}(\la T)} \partial_{r}^{2}\right)\Phi=-2\lambda^{2}\Phi\,.
\feq
On the spacetime boundaries $\cal{I^{-}}$ and $\cal{I^{+}}$ we obtain
again Eqs.\ (\ref{dilaton1}) and (\ref{dilaton2}), respectively. Using
Eqs.\ (\ref{Pinf}) and (\ref{had5}) of the Appendix, the two-point
boundary correlator is
\beq\lb{corr4}
\langle{\cal O}_{\phi}(r){\cal O}_{\phi}(r')\rangle= {\ka_{+} \over
\sin^{4}{a\la\over 2}(r-r')}\,,
\feq
where $r,r'\in \cal I^{-}$ or $r,r' \in\cal I^{+}$. Equation (\ref{corr4}) is
the correlator for an operator of conformal dimension $h=2$. In the dS$_{+}$
case we must also consider correlators between points $r,r'$, where $r\in{\cal
I^{-}}$ and $r'\in { \cal I^{+}}$. This corresponds to let $T\to-\infty$ and
$T'\to\infty$ in Eq.\ (\ref{int}). The function $P$ defined in Eq.\ (\ref{P})
satisfies the equation
\beq\lb{rifl}
P(T,r,T',r')= -  P(T,r,-T',r'+\pi)\,.
\feq
Using Eq.\ (\ref{rifl}) for $r\in {\cal I^{-}}$ and $r'\in {\cal I^{+}}$, 
${\cal J}$
becomes
\beq\lb{int2}
{\cal J}=\ka_{+}' \int drdr'\phi_{-1}(r)
\phi_{-1}(r'+\pi){1\over\sin^{4}{a\la\over 2}(r-r')}\,.
\feq
Analogously to the three-dimensional case of Ref.\ \cite{Strominger:2001pn}, we
can define the inverted boundary field $\tilde \phi_{-1}(r)=\phi_{-1}(r+\pi)$
and find the non-trivial correlations between points of the two different
boundaries. The correlator of boundary operators dual to $\tilde \phi_{-1}(r')$
and $\phi_{-1}(r)$ coincide with the two-point correlator on a single
boundary (\ref{corr4}). 
\section{Conclusions}
In this paper we have investigated the 2D dS/CFT correspondence.
The de Sitter conserved mass has been defined by exploiting a peculiar feature
of 2D dilaton gravity, namely the existence of a locally defined,
general covariant conserved charge. The dS$_{2}$/CFT$_{1}$ duality has been
implemented in analogy with the AdS$_{2}$/CFT$_{1}$ correspondence. We have
shown that the group of the asymptotic symmetries of dS$_{2}$ is equal to that
of AdS$_{2}$ and is generated by the same Virasoro algebra. The statistical
entropy of the de Sitter cosmological horizon coincides with the statistical
entropy of the AdS black hole. These results follow from the interpretation of
de Sitter spacetime as ``Wick rotated'' AdS spacetime. Similar conclusions have
been obtained for higher-dimensional de Sitter spacetime in Refs.\
\cite{Strominger:2001pn,Ghezelbash:2001vs,Myung:2001ab,Klemm:2001ea}. A major
difference between 2D de Sitter spacetime and the
higher-dimensional cases is the absence, in the former, of an upper bound for
the entropy. The entropy (\ref{entropy}) grows without limit with $a$. The
origin of the difference can be understood by comparing de Sitter in two and
three dimensions. The behavior of the entropy as a function of $a$ in $d=2$ and
$d=3$ is identical. However, the presence of the conical singularity in $d=3$
provides the upper bound $a=1$. For $d=2$ the spacelike coordinate $r$ is not a
radial coordinate. Therefore,  no ``natural'' normalization can be imposed on
it. In Ref.\ \cite{Cadoni:2002rr} it was argued that this feature follows from
the symmetry under dilatations of the model and is related with the 
impossibility of
establishing an area law in two dimensions.

In the second part of the paper we have computed the boundary correlators
of the model. We have found that the dS/CFT correspondence leads to
boundary operators of positive conformal dimension for 2D bulk tachyonic
perturbations. Non-causal tachyonic propagation in the bulk are usually
expected to lead to boundary operators with negative dimension, i.e., to
correlators which usually describe a non-unitary CFT, such as
${\langle\cal O}(r){\cal O}(r')\rangle\sim (r-r')^{l}$, where $l>0$. The
positivity of the conformal dimension seems to be strongly related to the
holographic nature of the dS/CFT correspondence. It would be very
interesting to understand whether this property is a peculiarity of the 2D
case or a general feature of the dS/CFT duality.
\section*{Appendix}
In this appendix we derive the 2D, de Sitter-invariant, Hadamard
two-point function for the dilaton field
\beq\lb{had}
G(X,X')=\langle 0|\{\Phi(X),\Phi(X')\}|0\rangle\,.
\feq
Equation (\ref{had}) is solution of the equation
\beq\lb{had1}
(\nabla^{2}_{X}-m^{2})G(X,X')=0\,,
\feq
where $m^{2}=-2\la^{2}$. Analogously to higher-dimensional cases, the $SO(1,1)$
invariant Green function can only be a function of the geodesic distance
$d(X,X')$, where $X,X'$ are coordinates of the three-dimensional Minkowski
embedding spacetime. It is convenient to introduce the quantity 
\beq\lb{geo}
P(X,X')=\la^{2}X^{A}X'^{B}\eta_{AB}\,,
\feq
where $P=\cos\la d$ and $\eta_{AB}=(1,1,-1)$ is the metric of the
three-dimensional Minkowski spacetime. Using Eq.\ (\ref{hyp}) $P$ can be
calculated for the three different parametrizations of dS$_{2}$. We 
have 
\eqn\lb{para}
{\rm dS}_0\;:&& \la X= e^{-\la T}\la r,\,\, \la Y=\cosh\la T -e^{\la T}(\la
r)^{2}/2,\,\,
\la Z=-\sinh\la T +e^{-\la T}(\la r)^{2}/2\,,\nonumber\\
{\rm dS}_{-}:&& \la X= \cosh\la T\,,\quad \la Y =\sinh\la T \sinh\la
r\,,\quad
\la Z =\sinh\la T \cosh\la r\,,\\
{\rm dS}_{+}:&& \la X =\cosh\la T\sin\la r\,,\quad \la Y= \cosh\la T 
\cos\la r\,,\quad
\la Z= \sinh\la T\,.\nonumber
\feqn
Substituting Eqs.\ (\ref{para}) in Eq.\ (\ref{geo}) we find
\eqn\lb{P}
{\rm dS}_0\,:&& P= \cosh\la(T-T')-\la^{2}e^{-\la
(T+T')}\left(r-r'\right)^{2}/2\,,\nonumber\\
{\rm dS}_{-}:&& P=
\cosh\la T\cosh\la T'-\sinh\la(T)\sinh\la T'\cosh\la(r-r')\,,\nonumber\\
{\rm dS}_{+}:&& P= \cosh\la T \cosh\la T'\cos\la(r-r')-\sinh\la T\sinh\la 
T'\,.
\feqn
The asymptotic behavior of $P$ at $T\to -\infty$ is
\eqn\lb{Pinf}
{\rm dS}_0\,:&& \lim _{T,T'\to -\infty} P= -{\la^{2}\over 2}e^{-\la
(T+T')}\left(r-r'\right)^{2}\,,\nonumber\\
{\rm dS}_{-}:&& \lim _{T,T'\to -\infty}P=-{1\over 2}e^{-\la
(T+T')}\sinh^{2} \la{(r-r')\over 2}\,,\nonumber\\
{\rm dS}_{+}:&& \lim _{T,T'\to -\infty}P=-{1\over 2}e^{-\la
(T+T')}\sin^{2} \la{(r-r')\over 2}\,.
\feqn
The behavior of $P$ on ${\cal I^{+}}$ can be obtained by changing the signs of
$T$ and $T'$ in Eqs.\ (\ref{Pinf}). For a generic scalar field $\phi$ of mass
$m$ propagating in 2D de Sitter spacetime, Eq.\ (\ref{had1}) is
\cite{Spradlin:2001pw, Candelas:du}
\beq\lb{had2}
\left[(1-P^{2}){d^{2}\over dP^{2}}-2P{d\over dP}- {m^{2}\over
\la^{2}}\right]G(P)=0\,.
\feq
The solution of Eq.\ (\ref{had2}) is
\beq\lb{had2b}
G={\rm Re}\,\rF(h_{+}, h_{-}, 1,z)\,,
\feq
where $\rF$ is the hypergeometric function, $z=(1+P)/2$, and $h_{\pm}= (1\pm
\sqrt{1-4m^{2}/\la^{2}})/2$. For the dilaton field  ($m^{2}=-2\la^{2}$)
the general solution of Eq. (\ref{had2}) can be expressed  in terms of 
elementary functions:
\beq\lb{had3}
G(P)=c_1\left(2-P\ln\left|{P+1\over P-1}\right|\right)+c_{2} P\,,
\feq
where $c_1,c_{2}$ are integration constants. The Green function $G$ is the sum
of two independent terms. The second term grows linearly with $P$ and is
singular in the $P\to\infty$ limit. This leads to divergences in Eq.\
(\ref{int}). The singularity can be removed by imposing the boundary condition
$c_{2}=0$. (It is not clear whether other physically acceptable boundary
conditions exist). The first term in Eq.\ (\ref{had3}) describes a Green
function with two singularities at $P=\pm 1$. The behavior at short distances
is that of a scalar field in a 2D spacetime. Near $P=1$, i.e., at
a geodesic distance $d=0$, $G$ is
\beq\lb{had4}
\lim_{P\to 1} G= 2c_1\ln(\la d)\,.
\feq
The integration constant $c_1$ can be determined by comparing Eq.\ (\ref{had4})
to the usual short-distance behavior in two-dimensions $G=-1/2\pi \ln (\la d)$.
The behavior at $P\to\infty$ is
\beq\lb{had5}
G(P)= -{2\over 3}{c_1\over P^{2}} +O(P^{-4})\,.
\feq
Finally, the ($c_2=0$) Green function is an even function of $P$.


\begin{thebibliography}{99}

\bibitem{Strominger:2001pn}
A.~Strominger,
JHEP {\bf 0110}, 034 (2001)
[arXiv:hep-th/0106113].

\bibitem{Spradlin:2001pw}
M.~Spradlin, A.~Strominger and A.~Volovich,
arXiv:hep-th/0110007.

\bibitem{Church:2001uy}
S.~Church, A.~Jaffe and L.~Knox,
arXiv:astro-ph/0111203.

\bibitem{Albrecht:2001xp}
A.~Albrecht, J.~A.~Frieman and M.~Trodden,
in {\it Proc. of the APS/DPF/DPB Summer Study on the Future of Particle Physics (Snowmass 2001) } ed. R.~Davidson and C.~Quigg,
arXiv:hep-ph/0111080.

\bibitem{Gibbons:mu}
G.~W.~Gibbons and S.~W.~Hawking,
Phys.\ Rev.\ D {\bf 15}, 2738 (1977).

\bibitem{Verlinde:2000wg}
E.~Verlinde,
arXiv:hep-th/0008140.

\bibitem{Cadoni:2002rr}
M.~Cadoni, P.~Carta and S.~Mignemi,
arXiv:hep-th/0202180.

\bibitem{Balasubramanian:2001nb}
V.~Balasubramanian, J.~de Boer and D.~Minic,
arXiv:hep-th/0110108.

\bibitem{Ghezelbash:2001vs}
A.~M.~Ghezelbash and R.~B.~Mann,
JHEP {\bf 0201}, 005 (2002)
[arXiv:hep-th/0111217].

\bibitem{Klemm:2001ea}
D.~Klemm,
Nucl.\ Phys.\ B {\bf 625}, 295 (2002)
[arXiv:hep-th/0106247].

\bibitem{Maldacena:1998ih}
J.~M.~Maldacena and A.~Strominger,
JHEP {\bf 9802}, 014 (1998)
[arXiv:gr-qc/9801096].

\bibitem{Myung:2001ab}
Y.~S.~Myung,
Mod.\ Phys.\ Lett.\ A {\bf 16}, 2353 (2001)
[arXiv:hep-th/0110123].

\bibitem{Ness:2002qr}
S.~Ness and G.~Siopsis,
arXiv:hep-th/0202096.

\bibitem{Cadoni:1999ja}
M.~Cadoni and S.~Mignemi,
Nucl.\ Phys.\ B {\bf 557}, 165 (1999)
[arXiv:hep-th/9902040].

\bibitem{Cadoni:2000fq}
M.~Cadoni and M.~Cavagli\`a,
Phys.\ Rev.\ D {\bf 63}, 084024 (2001)
[arXiv:hep-th/0008084].

\bibitem{Cadoni:2000kr}
M.~Cadoni and M.~Cavagli\`a,
Phys.\ Lett.\ B {\bf 499}, 315 (2001)
[arXiv:hep-th/0005179].

\bibitem{Cadoni:1994uf}
M.~Cadoni and S.~Mignemi,
Phys.\ Rev.\ D {\bf 51}, 4319 (1995)
[arXiv:hep-th/9410041].

\bibitem{Mann:1992yv}
R.~B.~Mann,
Phys.\ Rev.\ D {\bf 47}, 4438 (1993)
[arXiv:hep-th/9206044].
%
\bibitem{Carlip:1999cy}
S.~Carlip,
Class.\ Quant.\ Grav.\  {\bf 16}, 3327 (1999)
[arXiv:gr-qc/9906126].
%
\bibitem{Cadoni:2000ah}
M.~Cadoni and S.~Mignemi,
Phys.\ Lett.\ B {\bf 490}, 131 (2000)
[arXiv:hep-th/0002256].

\bibitem{Cadoni:2000gm}
M.~Cadoni, P.~Carta, D.~Klemm and S.~Mignemi,
Phys.\ Rev.\ D {\bf 63}, 125021 (2001)
[arXiv:hep-th/0009185].

\bibitem{Yo}
J. W. York, Found. Phys. {\bf16}, 249 (1986).

\bibitem{Candelas:du}
P.~Candelas and D.~J.~Raine,
Phys.\ Rev.\ D {\bf 12}, 965 (1975).
\end{thebibliography}
\end{document}